\documentclass[aps,reprint,groupedaddress,superscriptaddress, amsmath,amssymb]{revtex4-1}
\usepackage{amsmath,amssymb,bm,mathrsfs,graphicx,braket,amsthm,enumerate,times,color,here,comment}
\usepackage[colorlinks=true,citecolor=blue,linkcolor=blue]{hyperref}

\begin{document} 

\title{Connecting higher-order topological insulators to lower-dimensional topological insulators}

\author{Akishi Matsugatani}
\affiliation{Department of Applied Physics, The University of Tokyo, Hongo, Bunkyo-ku, Tokyo 113-8656, Japan}

\author{Haruki Watanabe}
\email[]{haruki.watanabe@ap.t.u-tokyo.ac.jp}
\affiliation{Department of Applied Physics, The University of Tokyo, Hongo, Bunkyo-ku, Tokyo 113-8656, Japan}

\date{\today}

\begin{abstract}
In recent years, the role of crystal symmetries in enriching the variety of TIs have been actively investigated.  
Higher-order TIs are a new type of topological crystalline insulators that exhibit gapless boundary states whose dimensionality is lower than those on the surface of conventional TIs.   In this paper, relying on a concrete tight-binding model, we show that higher-order TIs can be smoothly connected to conventional TIs in a lower dimension without the bulk-gap closing or symmetry breaking. Our result supports the understanding of higher-order TIs as a stacking of lower-dimensional TIs in a way respecting all the crystalline symmetry.
\end{abstract}

\maketitle

\section{Introduction}\label{sec:intro}

\begin{figure*}[t]
\begin{center}
\includegraphics[clip,width=0.6\textwidth]{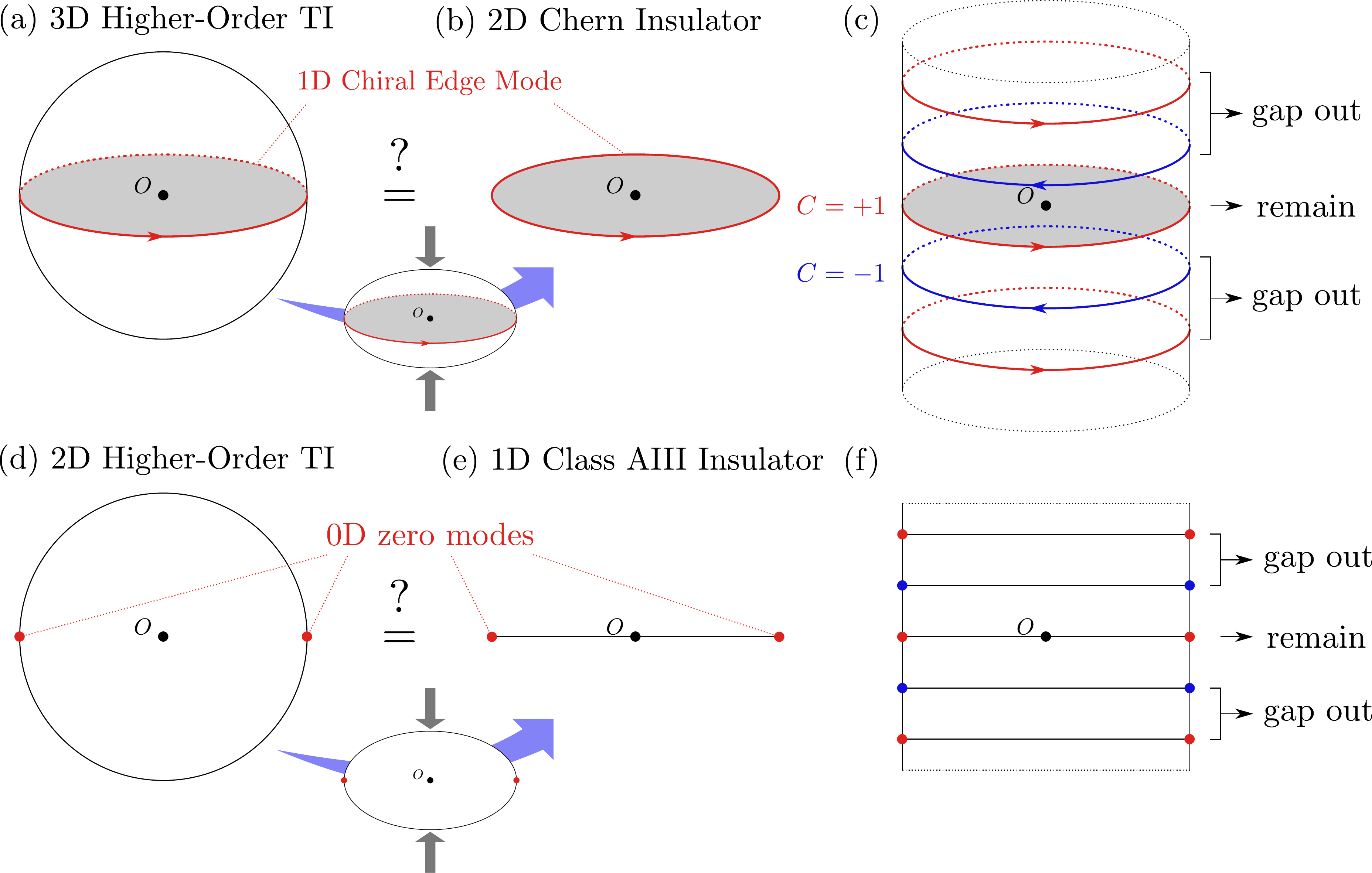}
\caption{A schematic illustration of 
(a) a 3D HOTI, 
(b) a 2D Chern insulator, 
(c) a staggered stacking of Chern insulators,
(d) a 2D HOTI, 
(e) a 1D TI in class AIII,
and (f) a staggered stacking of 1D TI in class AIII. $O$ is the inversion center.  Red (blue) circles represent a right-moving (left-moving) chiral edge mode, while red (blue) dots represent a zero-energy edge mode corresponding to the index $+1$ $(-1)$. }
\label{fig:hoti-ci}
\end{center}
\end{figure*}

Recently a new class of topological crystalline insulators, the so-called higher-order topological insulators (HOTIs) have attracted growing research interest~\cite{SitteRoschAltmanFritz, ZhangKaneMele,Teo,  Benalcazar17,  Hashimoto2017, Song17, Schindler17,  Benalcazar17Aug,Fang17, Langbehn17,Eslam1,Eslam2,Brouwer,Ezawa1,Bismuth,Ezawa2,Ezawa3}. Their defining feature is, like the standard topological insulators (TIs), the appearance of gapless boundary states, but, unlike the TIs, the dimensionality of the gapless modes is reduced to less than $d-1$ for a $d$-dimensional ($d$D) insulating bulk ($d=1,2,3,\ldots$). For example, Fig.~\ref{fig:hoti-ci} (a) illustrates the case of inversion symmetric 3D HOTI with broken time-reversal symmetry, which hosts an equatorial 1D chiral edge mode on the surface~\cite{SitteRoschAltmanFritz, ZhangKaneMele}.  Although the fundamental group $\pi_1$ of a sphere is trivial, there is no way to shrink the 1D ring to a point without breaking the assumed inversion symmetry. Depending on the specific symmetry settings and dimensions, there is a variety of HOTIs, exhibiting 0D corner states [e.g. Fig.~\ref{fig:hoti-ci} (d)] or 1D helical edge modes [e.g. combination of Fig.~\ref{fig:hoti-ci} (a) and its time-reversal copy], for example.

The chiral edge mode in Fig.~\ref{fig:hoti-ci} (a) is the reminiscent of the 2D Chern insulator with the unit Chern number.  Then a fundamental question arises: Is the physical property of the 3D HOTI essentially the same as that of the 2D Chern insulator? In other words, are these two insulators smoothly connected to each other without closing the bulk gap or breaking the symmetry, despite the apparent difference in their dimensionality? One can imagine squishing the ball in Fig.~\ref{fig:hoti-ci} (a) into the disk in Fig.~\ref{fig:hoti-ci} (b) in such a way that the equator of the ball coincides with the boundary of the disk, as illustrated by the blue arrow in between Fig.~\ref{fig:hoti-ci} (a,b), and see if the bulk excitation gap and the assumed symmetries are preserved during the process.  Similarly, one can ask if the 2D HOTI with 0D zero-energy modes can be smoothly deformed into a 1D TI with the same edge states.  In this paper, we explicitly perform this analysis using a cube, a square, and a line segment instead, since these are much easier to handle on a lattice. 

The possible equivalence between the 3D HOTIs and the 2D Chern insulators and that between the 2D HOTIs and the 1D TIs are just a particular instance of more general connection between HOTIs with TIs in lower dimensions. Depending on the concrete symmetry settings, the Chern insulator should be replaced with an appropriate 2D topological phase.  In fact, the coupled-layer construction of HOTIs proposed in Refs.~\cite{HaoSong,ShengJieHuang,Fang17} supports their equivalence. For example, one can form a staggered stacking of Chern insulators with $C=\pm1$ in an inversion symmetric manner. Suppose that the inversion center is included in one of the Chern insulators with $C=+1$. Then, keeping the inversion symmetry, gapless chiral edge modes with opposite chirality can be gapped out in pairs as illustrated in Fig.~\ref{fig:hoti-ci} (c), leaving only the single chiral mode. The 3D insulator constructed as such gives rise to a particular instance of a HOTI protected by the inversion symmetry. Since the $z=0$ layer is completely decoupled from others, at least the low-energy property, insensitive to gapped layers in $z\neq0$, must be identical to the 2D Chern insulator. However, this observation should not be taken as the general proof of the equivalence, since, in principle, there can be a HOTI that cannot be constructed via the coupled-layer construction. (In fact, searching for a concrete case of such HOTIs would also be a subject of an independent study.)

In this paper, we provide an alternative argument that connects HOTIs to a lower-dimensional TIs without relying on the coupled layer construction. Our analysis is based on two concrete tight-binding models, whose thickness will be systematically controlled to bridge the $d$D and $d-1$D limit for $d=3$ and $2$.  This is just a model-dependent argument and hence is no more general than the above understanding via the coupled layer construction.  However, we still believe the explicit analysis presented below helps to grasp the generic nature of HOTIs.

\section{3D HOTI to 2D Chern insulator}
\subsection{Tight-binding model for a topological insulator}
\label{sec:ti-mag}
In order to construct a concrete tight-binding model of a HOTI with chiral hinge modes, we start from a time-reversal invariant 3D topological insulator, which has gapless Dirac surface states on each surface. Let us take a four-band tight-binding model~\cite{Qi2008}
\begin{equation}\label{eq:3D-ti-model}
H(\vec{k})=-t\sum_j\sin k_j\ \tau_x\otimes\sigma_j-(m-c\sum_j \cos k_j)\tau_z\otimes\sigma_0,
\end{equation}
where the sum of $j$ is over $j=x,y,z$, $\tau_j$ and $\sigma_j$ are Pauli matrices, and $\tau_0$ and $\sigma_0$ are the $2\times2$ identity matrix. To simplify the analysis we set $t=c=1$ in the following. This model has the inversion symmetry $I=\tau_z\otimes\sigma_0$ and the time-reversal symmetry $\mathcal{T}=-i\tau_0\otimes\sigma_y K$, where $K$ represents the complex conjugation. 

The value $m=2$ at the half filling falls into one of topological phases protected by the time-reversal symmetry. Figure~\ref{fig:3D-ti} (a) shows the inversion parity of two valence bands at each time-reversal invariant momentum. Only the $\Gamma$ point [$\vec{k}=(0,0,0)$] has the odd parity and others have the even parity. According to the Fu-Kane formula, this combination of the inversion parity implies the nontrivial strong index, while all weak indices vanish~\cite{FuKane}. To confirm the appearance of the surface Dirac dispersion we compute the band dispersion under the periodic boundary condition (PBC) in $x$ and $y$ and the open boundary condition (OBC) in $z$ with $L_z=2l_z+1=25$ layers, as shown in Fig.~\ref{fig:3D-ti} (b,c). As expected, bulk states have a band gap of the order 1 and the gapless surface states, localized near $z=\pm l_z$, have a Dirac-like linear dispersion around $(k_x, k_y) = (0,0)$.

\subsection{Weak uniform magnetic field}
When the time-reversal symmetry is explicitly broken by perturbations, the Dirac surface modes may acquire a mass gap without closing the bulk gap.  Below we discuss the effect of a uniform magnetic field $\vec{B}=B(0,-\sin\theta,\cos\theta)$ by adding a term $-\tau_0\otimes\vec{B}\cdot\vec{\sigma}$ everywhere both on the surfaces and in the bulk.  Note that the inversion symmetry stays unbroken even in the presence of $\vec{B}$ and the bulk gap does not close as far as $\vec{B}$ is sufficiently small.

\begin{figure}[H]
\begin{center}
\includegraphics[clip,width=0.7\columnwidth]{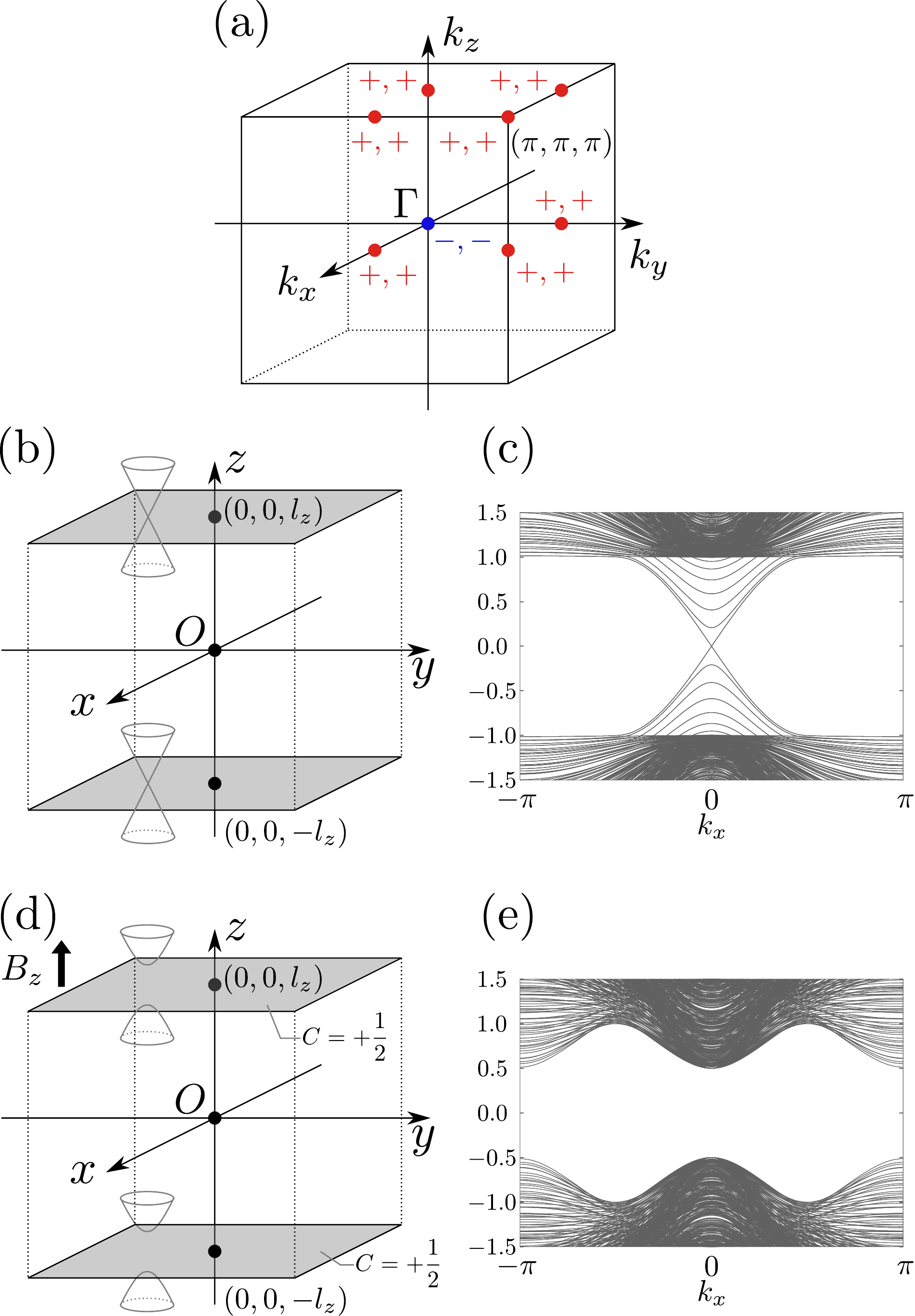}
\caption{(a) The parity eigenvalue of the valence bands of the model \eqref{eq:3D-ti-model} for $m=2$ and $t=c=1$. (b,c) The top and bottom surfaces host gapless Dirac states protected by the time-reversal symmetry. (d,e) A uniform magnetic field $\vec{B}=(0,0,\frac{1}{2})$ opens an excitation gap. In panels (c,e), the number of layers in $z$ is set to be $L_z=25$, and $30$ different values of $k_y$ are overlaid.}
\label{fig:3D-ti}
\end{center}
\end{figure}

\begin{figure*}
\begin{center}
\includegraphics[clip,width=0.8\textwidth]{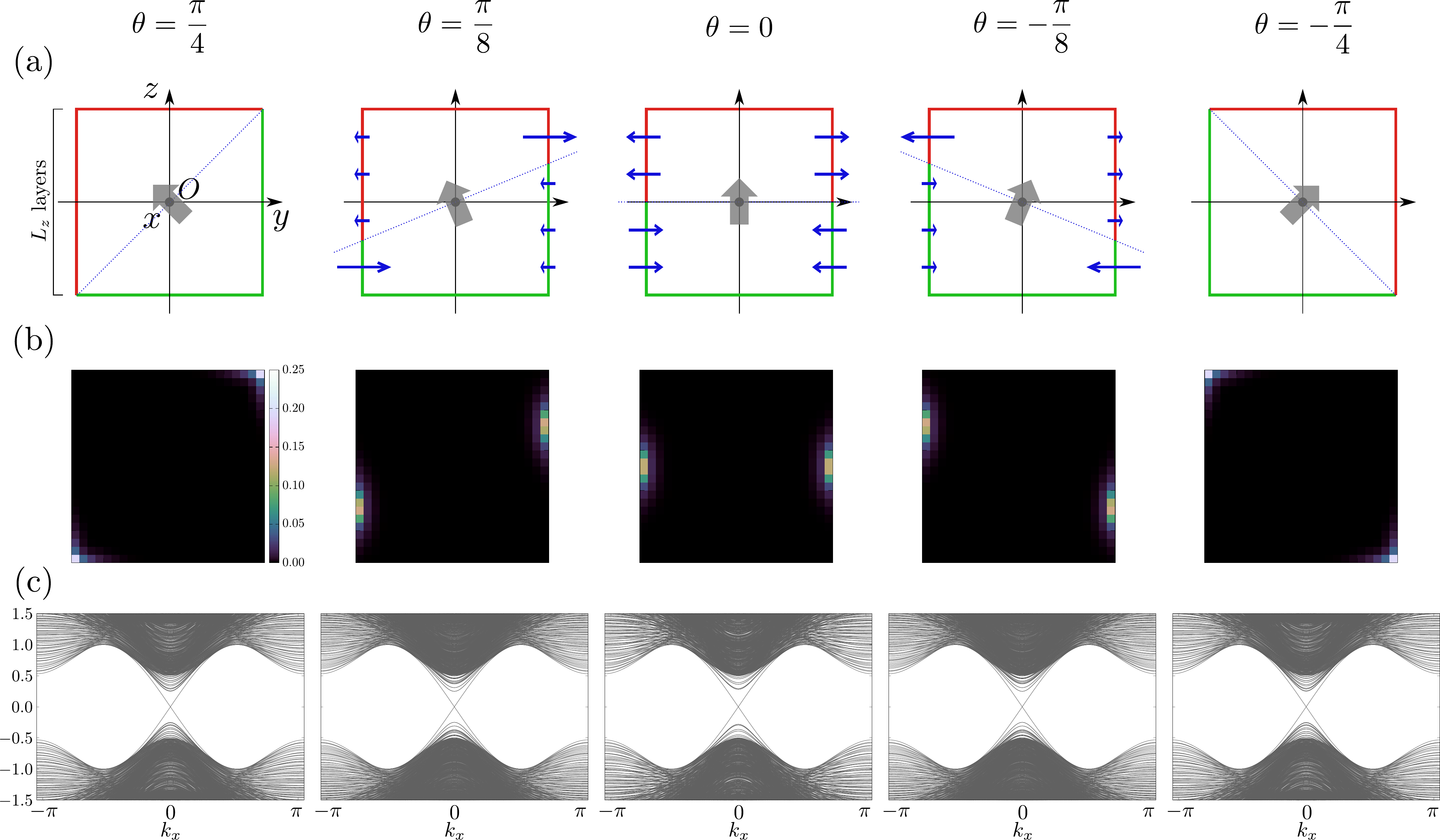}
\caption{3D TI under a uniform magnetic field $\vec{B}=\frac{1}{2}(0,-\sin\theta,\cos\theta)$ for $\theta=\frac{\pi}{4}$, $\frac{\pi}{8}$, $0$, $-\frac{\pi}{8}$, and $-\frac{\pi}{4}$.
For $|\theta|<\frac{\pi}{4}$, an additional surface Zeeman field [defined in Eqs.~\eqref{eq:def-mag-yz1} and \eqref{eq:def-mag-yz2}] is applied as illustrated in (a). The boundary condition in $x$ is PBC and that for $y$ and $z$ is OBC. The red and green color in (a) indicate the sign of the mass gap.  The panel (b) is the density plot of the weight of the zero-energy states in the band structure (c).}
\label{fig:rot_mag_3d}
\end{center}
\end{figure*}

As an example, Fig.~\ref{fig:3D-ti} (e) shows the energy dispersion for the case $\theta=0$ and $B=\frac{1}{2}$. The magnitude of the mass gap of the Dirac surface modes is proportional to the magnitude of the normal component  $B_\perp\equiv\vec{B}\cdot\vec{n}$ as long as $|B_\perp|$ is small.  (In this particular setting, when $|B_\perp|$ exceeds $\frac{1}{2}$, the gap at $k_z=\pi$ becomes smaller than that at $k_z=0$.)  Now that two-dimensional surfaces are completely gapped by the time-reversal breaking perturbation, one might think the band insulator became completely trivial. However, as long as the inversion symmetry is respected, this is not the case --- in fact, the topological insulator under an external magnetic field can be regarded as a HOTI exhibiting a 1D chiral edge state on its surface \cite{SitteRoschAltmanFritz} and as we will see in Sec.~\ref{pi4hoti}.  This is hinted by the combination of the inversion parity.  Since the bulk gap did not close by the applied magnetic field, the resulting band insulator still has the same inversion parity as in Fig.~\ref{fig:3D-ti} (a). According to Refs.~\cite{NC,Ono}, this inversion parity falls into the class $(0,0,0,2)$ in the $\mathbb{Z}_2\times\mathbb{Z}_2\times\mathbb{Z}_2\times\mathbb{Z}_4$ classification for the space group $P\bar{1}$ generated by the inversion and the 3D lattice translation.

One way of understanding this nontrivial topology is through the bulk Chern number.  When the $z$-component of the magnetic field is positive, the gapped surfaces at the top and bottom can independently be thought of a Chern insulator with $C=+\frac{1}{2}$ as a single Dirac cone opened a gap on each surface. (Here, the normal direction is set to be $(0,0,1)$ for both the top and the bottom surface.) Consequently, the $2L_z$ bands in total below the Fermi energy $\mu=0$ (as a function of $k_x$ and $k_y$) in Fig.~\ref{fig:3D-ti} (e) has the net Chern number $C=+1$, which we confirmed explicitly using our tight-binding model.  This is in a sharp contrast to the \emph{vanishing} Chern number of the purely 3D insulating bulk in the model~\eqref{eq:3D-ti-model} under the PBC in every direction.  The non-zero Chern number is generated in the process of making a finite-thickness slab and opening a gap to its surfaces.  This Chern number implies the presence of protected gapless states on the side surface.

\subsection{Chiral hinge mode}
\label{pi4hoti}
To make a concrete connection to the HOTI with a chiral hinge mode, let us examine the surface property more carefully by taking the OBC in both $y$ and $z$ directions with $l_y=l_z$, while keeping the PBC in $x$. We set $\theta=\frac{\pi}{4}$ for the uniform magnetic field $\vec{B}=B(0,-\sin\theta,\cos\theta)$ ($B=\frac{1}{2}$) so that every 2D surface has a nonzero normal component $B_\perp$. See the illustration in the leftmost panel of Fig.~\ref{fig:rot_mag_3d} (a).  We find the appearance of gapless modes localized to the hinge $(y,z)=\pm(l_y,l_z)$ as demonstrated in Fig.~\ref{fig:rot_mag_3d} (b) and (c). This confirms that the TI under the magnetic field is indeed a HOTI at least for this choice of the uniform magnetic field.

\subsection{Moving the position of the chiral mode}
\label{moving}
As the first step toward addressing our original question on the equivalence of the 3D HOTI and a 2D Chern insulator, let us adiabatically rotate the direction of the external magnetic field from $\theta=\frac{\pi}{4}$ to $\theta=0$. This step is not at all inevitable, but makes it easier to reduce the number of layers in $z$ in Sec.~\ref{sec:3d-2d}.

If we natively set $\theta=0$, the side surfaces lose the normal component of the magnetic field and becomes entirely gapless. To avoid this, we introduce a term $-b_y(\vec{x}) \tau_0\otimes\sigma_y$ describing an additional Zeeman field normal to the side surfaces.  Here, $b_y(\vec{x})$ is applied only to the surface in such a way that
(i) the inversion symmetry and the translation symmetry in $x$ are preserved and 
(ii) the absolute value of the $y$-component of the total magnetic field on each side surface becomes independent of $\theta$. Specifically, 
\begin{eqnarray}\label{eq:def-mag-yz1} 
b_y(\vec{x}) =
\begin{cases}
B(\sin\theta+\frac{1}{\sqrt{2}}) & (l_z\tan\theta< z\leq l_z)\\
B\sin\theta & (z=l_z\tan\theta)\\
B(\sin\theta-\frac{1}{\sqrt{2}}) & (-l_z\leq z<l_z\tan\theta)
\end{cases}
\end{eqnarray}
on the surface $y=l_y$ and
\begin{eqnarray}\label{eq:def-mag-yz2} 
b_y(\vec{x}) =
\begin{cases}
B(\sin\theta+\frac{1}{\sqrt{2}}) & (-l_z\leq z<-l_z\tan\theta)\\
B\sin\theta & (z=-l_z\tan\theta)\\
B(\sin\theta-\frac{1}{\sqrt{2}}) & (-l_z\tan\theta< z\leq l_z)
\end{cases}
\end{eqnarray}
on the other surface $y=-l_y$, as illustrated in Fig.~\ref{fig:rot_mag_3d} (a).

Figure~\ref{fig:rot_mag_3d} (b) shows the density profile of the gapless modes for different values of $\theta$. Clearly they move smoothly as $\theta$ changes. In this process, the bulk gap does not close as demonstrated in Fig.~\ref{fig:rot_mag_3d} (c). The fact that $(d-2)$-dimensional states of HOTI can move away from hinge or corner is pointed out in Ref.~\cite{Song17}. They can be moved but cannot be removed without breaking the protecting symmetry (the inversion symmetry in our model).

\subsection{Bridging the 3D and the 2D limit}\label{sec:3d-2d}

\begin{figure}[t]
\begin{center}
\includegraphics[clip,width=0.5\textwidth]{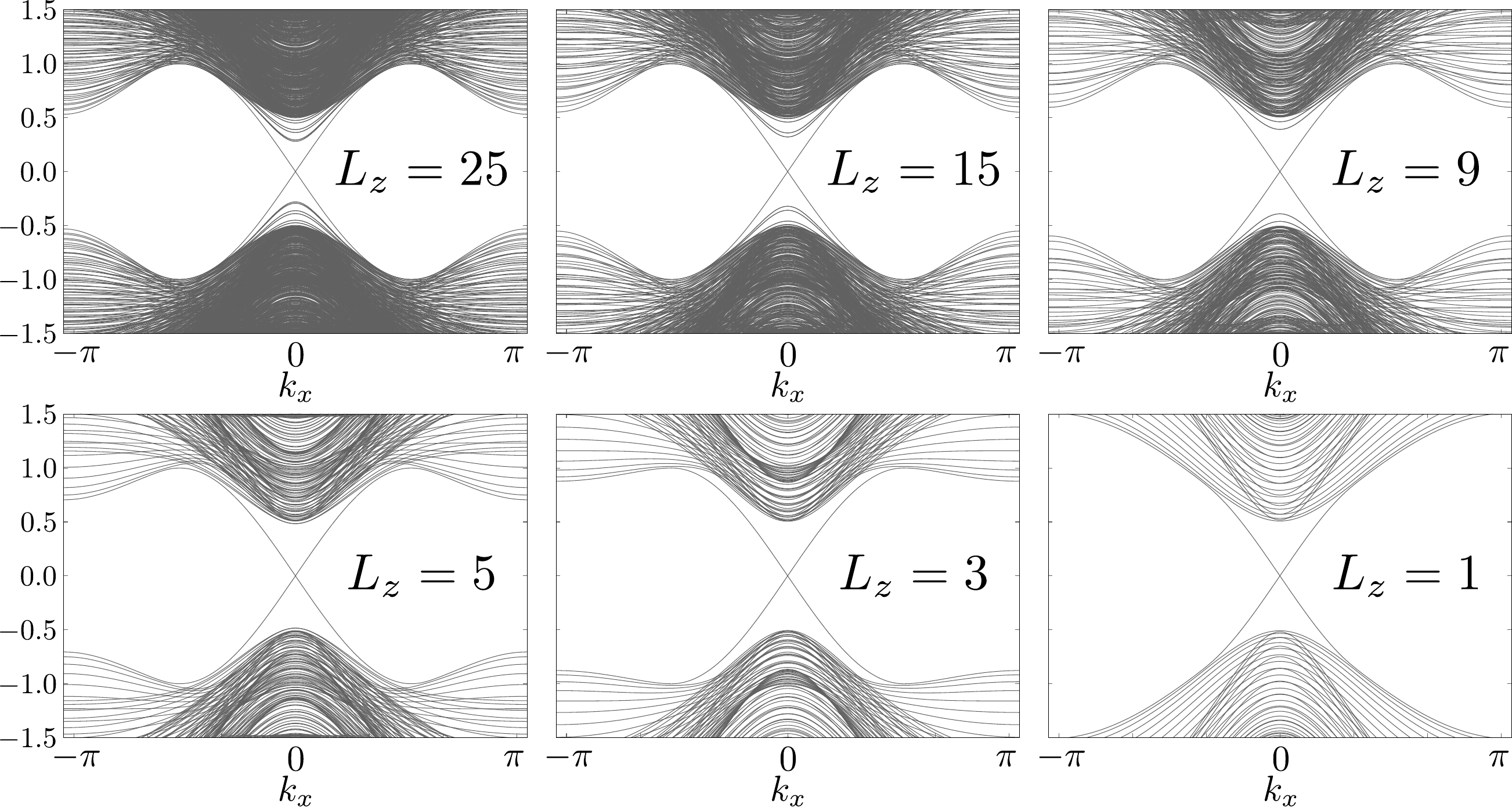}
\caption{The band dispersion computed under the OBC in $y$ and $z$ and PBC in $x$ for different number of layers: $L_z=25$, $15$, $9$, $5$, $3$, and $1$.}
\label{fig:3D-tci}
\end{center}
\end{figure}

We are ready address our original question whether the 3D HOTI can be adiabatically connected to the 2D Chern insulator.  To this end, we start from the $\theta=0$ case in the previous section and reduce the system size in the $z$ direction one by one as illustrated in Fig.~\ref{fig:continuous}.  Since we are treating a model on a lattice, the number of layers in $z$ can be reduced only discretely, but as far as the gapless edge modes and the band gap are concerned we find a smooth deformation as illustrated in Fig.~\ref{fig:3D-tci}. The smoothness in the deformation will be further checked by continuously reducing the coupling between $z=l_z$ and $z=l_z+1$ layers, instead of abruptly switching it off in Sec.~\ref{sec:smooth-deformation}. The only change occurred in the process of reducing the number of layers is that the band structure corresponding to the bulk states becomes more and more sparse. Most importantly, we did not observe any drastic change in the chiral edge mode and the bulk gap. 

The resulting single-layer insulator is described by the model in Eq.~\eqref{eq:3D-ti-model} (but the sum of $j$ is now restricted to $j=x$ and $y$) plus the magnetic field $-B_z \tau_0\otimes\sigma_z$:
\begin{eqnarray}\label{H0}
H_0(\vec{k})=&&-\sum_{j=x,y}\sin k_j\ \tau_x\otimes\sigma_j\notag\\
&&-(2-\sum_{j=x,y} \cos k_j)\tau_z\otimes\sigma_0-\frac{1}{2}\ \tau_0\otimes\sigma_z.
\end{eqnarray}
In this four band model, the two valence bands have the Chern number $+1$. The minimal model for a Chern insulator requires only one filled band (together with other unfilled bands), and we found no obstruction to induce an additional band gap between the bottom two bands in this model.  This concludes the adiabatic
deformation of the 3D HOTI to the most elementary Chern insulator.

\subsection{Smooth reduction of layers}\label{sec:smooth-deformation}
\begin{figure}
\begin{center}
\includegraphics[clip,width=1.05\columnwidth]{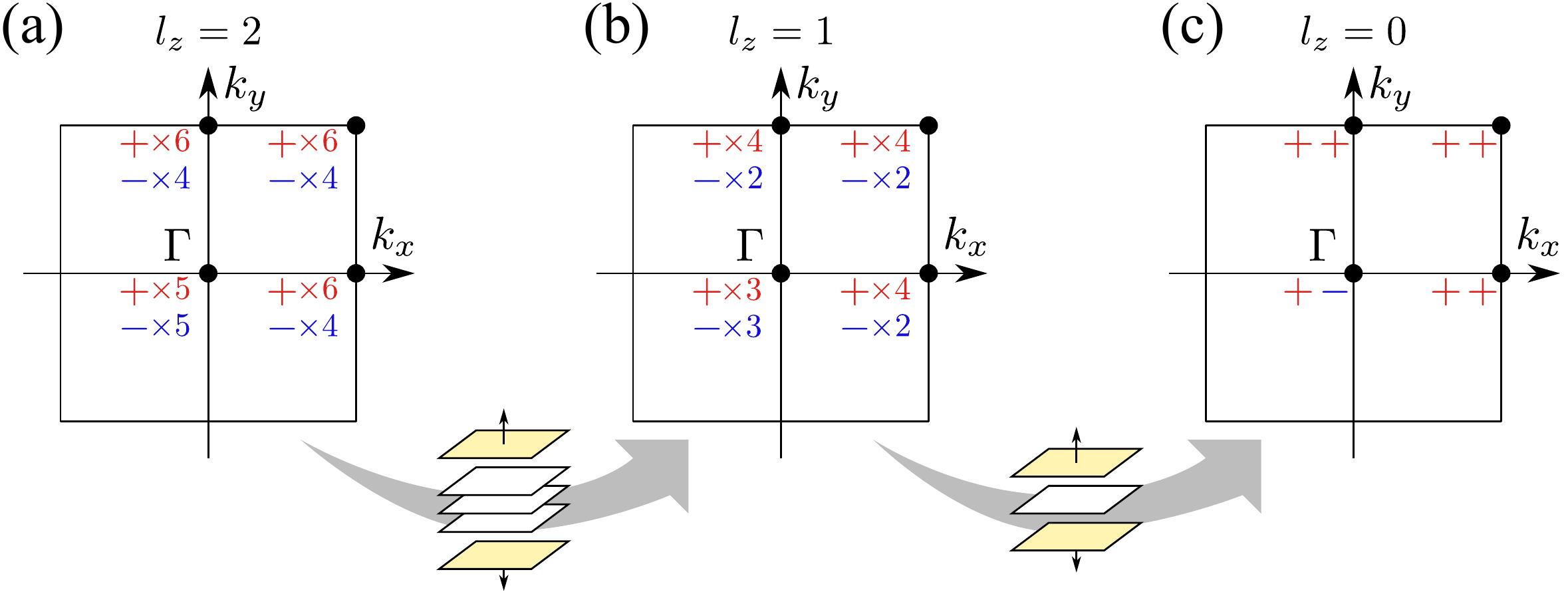}
\caption{The parity eigenvalue of the valence bands of the model in Eq.~\eqref{eq:3D-ti-model} for 5 layers (a), 3 layers (b), and 1 layer (c).}
\label{fig:continuous}
\end{center}
\end{figure}

In the previous section, the system size is reduced one by one. Here we perform a more smooth deformation by continuously switching off the inter-layer coupling.  As an example, we discuss reducing the number of layers from $2l_z+1=5$ to $2l_z+1=3$ as illustrated in Figs.~\ref{fig:continuous} (a) and (b). Each layer contains four bands as a 2D system and the full tight-binding model we discuss contains $4(2l_z+1)=20$ bands in total. 

The interpolating Hamiltonian reads
\begin{eqnarray}
H^{(l_z=2)}_\alpha&=&(1-\alpha)H^{(l_z=2)}_{0}+\alpha H^{(l_z=2)}_{1},\\
H^{(l_z=2)}_{0}&=&\left(\begin{matrix}H_0(\vec{k})& H_z & 0 & 0 & 0\\ H_z^\dagger & H_0(\vec{k}) & H_z & 0 & 0\\ 0 & H_z^\dagger & H_0(\vec{k}) & H_z & 0\\ 0 & 0 & H_z^\dagger & H_0(\vec{k}) & H_z\\ 0 & 0 & 0 & H_z^\dagger & H_0(\vec{k})\end{matrix}\right),\\
H^{(l_z=2)}_{1}&=&\left(\begin{matrix}H_I& 0 & 0 & 0 & 0\\ 0& H_0(\vec{k}) & H_z & 0 & 0\\ 0 & H_z^\dagger & H_0(\vec{k}) & H_z & 0\\ 0 & 0 & H_z^\dagger & H_0(\vec{k}) & 0\\ 0 & 0 & 0 & 0 & H_I\end{matrix}\right),
\end{eqnarray}
where $H_0(\vec{k})$ is given in Eq.~\eqref{H0},
\begin{eqnarray}
H_z=-\frac{1}{2}\ \tau_z\otimes\sigma_0+\frac{i}{2}\ \tau_x\otimes\sigma_z
\end{eqnarray}
is the inter-layer coupling originating from the nearest-neighbor hopping term in $z$ in Eq.~\eqref{eq:3D-ti-model}, and
\begin{eqnarray*}
H_I=-2\ \tau_z\otimes\sigma_0
\end{eqnarray*}
describes the trivial layers after decoupling. The layer at $z=j$ ($-2\leq j\leq 2$) corresponds to the $(3-z)$-th block in $H^{(l_z=2)}_\alpha$.  The inversion symmetry $I$ is implemented as
\begin{equation*}
I=\left(\begin{matrix} 0 & 0 & 0 & 0 & \tau_z\otimes\sigma_0\\ 0 & 0 & 0 & \tau_z\otimes\sigma_0 & 0\\ 0 & 0 & \tau_z\otimes\sigma_0 & 0 & 0\\ 0 & \tau_z\otimes\sigma_0 & 0 & 0 & 0\\ \tau_z\otimes\sigma_0 & 0 & 0 & 0 & 0\end{matrix}\right),
\end{equation*}
satisfying $IH(\alpha;\vec{k})=H(\alpha;-\vec{k})I$.

The parameter $\alpha\in[0,1]$ interpolates the original 5 layer system [Fig.~\ref{fig:continuous} (a)] and the 3 layer system together with the two decoupled 2 layers at $z=\pm2$ [Fig.~\ref{fig:continuous} (b)].  The band gap remains almost unaffected as $\alpha$ is increased from $0$ to $1$. We also performed the same analysis connecting $2l_z+1=3$ layers to $2l_z+1=1$ layer [Fig.~\ref{fig:continuous} (c)] and the result was the same.  These results support the discussion in Sec.~\ref{sec:3d-2d} that the 3D HOTI can be smoothly connected to the Chern insulator.

\section{2D HOTI to 1D topological insulator}
\label{sec:2d-case}
Now we move on to the relation between the 2D HOTI and the 1D TI.  Our discussion follows completely the same steps as in the previous section.

We start from the time-reversal invariant 2D topological insulator with 1D helical edge states. We reuse the same tight-binding model \eqref{eq:3D-ti-model} but in the 2D limit. The sum of $j$ is hence restricted to $j=x$ and $y$.  In addition to the inversion symmetry $I=\tau_z\otimes\sigma_0$ and the time-reversal symmetry $\mathcal{T}=-i\tau_0\otimes\sigma_y K$, the 2D model has the chiral symmetry $\Pi=\tau_x\otimes\sigma_z$. The full internal symmetry of this model is thus class DIII of the Altland-Zirnbauer symmetry classes.

\begin{figure}[t]
\begin{center}
\includegraphics[clip,width=0.7\columnwidth]{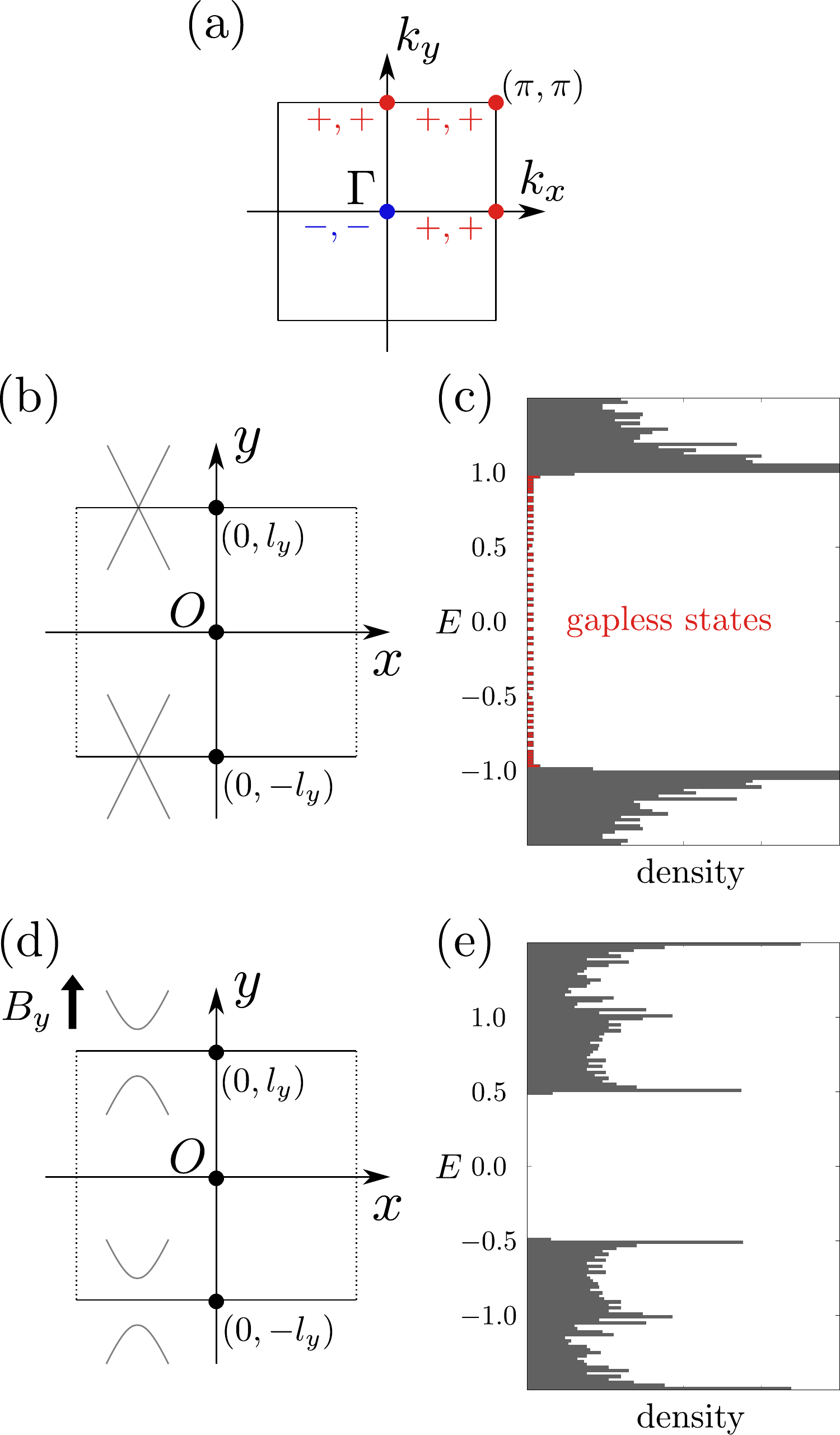}
\caption{(a) The parity eigenvalue of the valence bands of the model in Eq.~\eqref{eq:3D-ti-model} for $t=c=m=1$. (b,c) Two edges at $y=\pm l_y$ host a helical edge state protected by the time-reversal symmetry. (d,e) The uniform magnetic field $\vec{B}=(0,\frac{1}{2},0)$ opens a excitation gap. In panels (c,e), the number of layers in $y$ is set to be $L_y=2l_y+1=25$ and $120$ different values of $k_x$ are overlaid.}
\label{fig:2D-ti}
\end{center}
\end{figure}

The inversion parities of two valence bands for the choice of parameters $t=c=m=1$ are shown in Fig.~\ref{fig:2D-ti}. This combination of parity eigenvalues implies the nontrivial $\mathbb{Z}_2$ quantum spin Hall index~\cite{FuKane} and thus helical edge states protected by the time-reversal symmetry should appear. Figure~\ref{fig:2D-ti} (c) shows the density of states under the PBC in $x$ and the OBC in $y$ with $L_y=2l_y+1=25$ layers illustrated in Fig.~\ref{fig:2D-ti} (b). We indeed observe the in-gap states that are localized around $y=\pm l_y$.

Next we break the time-reversal symmetry by applying a uniform magnetic field $\vec{B}=B(-\sin\theta,\cos\theta,0)$ with 
$\theta=0$ and $B=\frac{1}{2}$ as shown in Fig.~\ref{fig:2D-ti} (d). The corresponding density of states is shown in Fig.~\ref{fig:2D-ti} (e).  As expected, the edge state acquires a mass gap of the order of $|B_\perp|$.

\begin{figure*}
\begin{center}
\includegraphics[clip,width=0.8\textwidth]{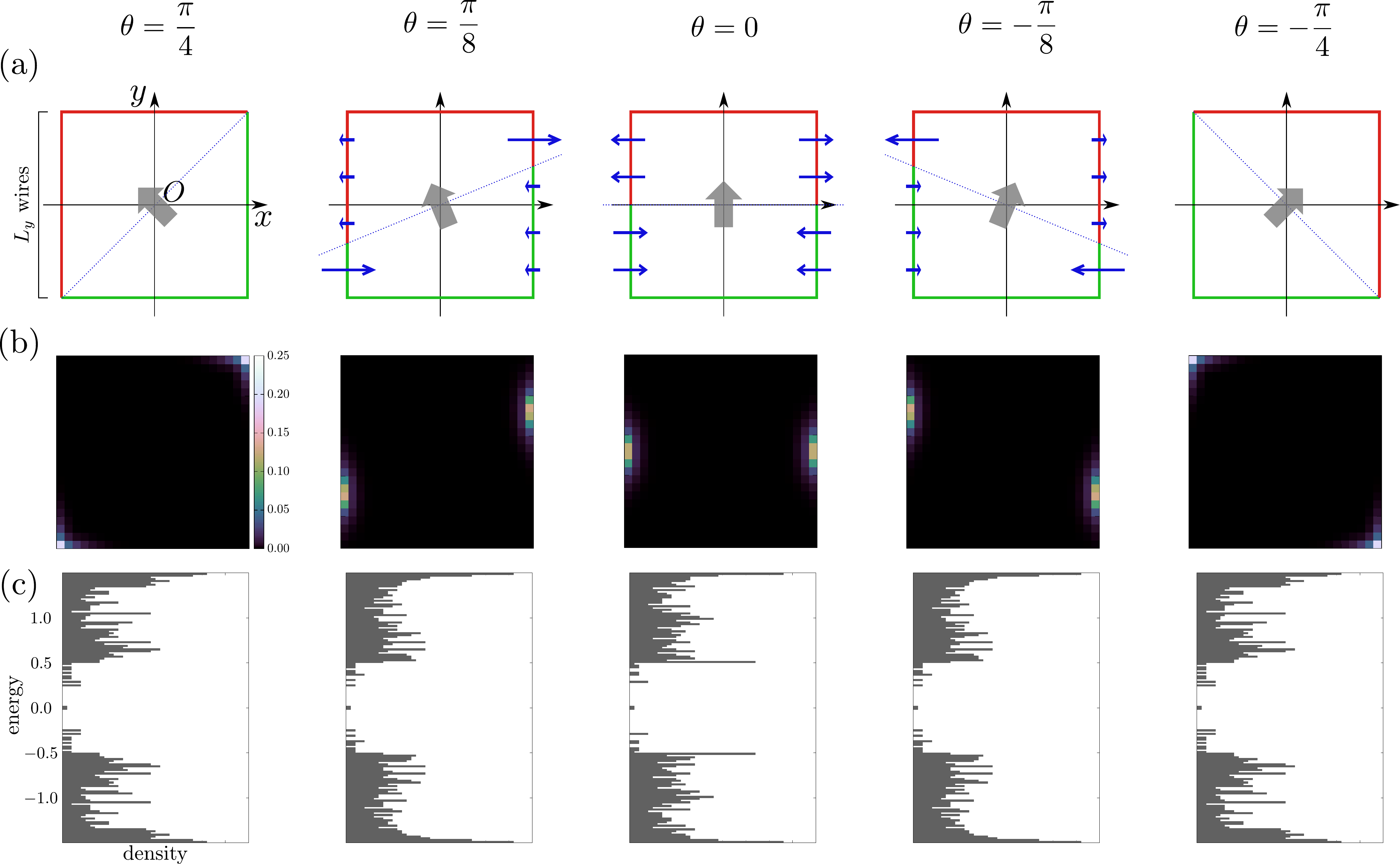}
\caption{2D TI with chiral symmetry under a uniform mangetic field $\vec{B}=B(-\sin\theta,\cos\theta,0)$ for $\theta=\frac{\pi}{4}$, $\frac{\pi}{8}$, $0$, $-\frac{\pi}{8}$, and $-\frac{\pi}{4}$. An additional Zeeman field [defined in Eqs.~\eqref{eq:def-mag-xy1} and \eqref{eq:def-mag-xy2}] is applied as illustrated in (a). The boundary condition is  OBC in both $x$ and $y$ directions. The panel (b) is the density plot of the weight of the zero-energy modes in the panel (c).}
\label{fig:rot_mag_2d}
\end{center}
\end{figure*}

Since the chiral symmetry remains unbroken, our 2D model under the magnetic field belongs to the class AIII. When a suitable boundary condition is taken, it exhibits 0D zero-energy modes protected both by the inversion symmetry and the chiral symmetry.  To see this, we take the OBC both in $x$ and $y$ direction.  We introduce the uniform magnetic field $\vec{B}=B(-\sin\theta,\cos\theta,0)$ together with the Zeeman field $-b_x(\vec{x})\tau_0\otimes\sigma_x$ localized to the side edges. We set
\begin{eqnarray}\label{eq:def-mag-xy1} 
b_x(\vec{x}) =
\begin{cases}
B(\sin\theta+\frac{1}{\sqrt{2}}) & (l_y\tan\theta< y\leq l_y)\\
B\sin\theta & (y=l_y\tan\theta)\\
B(\sin\theta-\frac{1}{\sqrt{2}}) & (-l_y\leq y<l_y\tan\theta)
\end{cases}
\end{eqnarray}
on the edge $x=l_x$ and
\begin{eqnarray}\label{eq:def-mag-xy2} 
b_x(\vec{x}) =
\begin{cases}
B(\sin\theta+\frac{1}{\sqrt{2}}) & (-l_y\leq y<-l_y\tan\theta)\\
B\sin\theta & (y=-l_y\tan\theta)\\
B(\sin\theta-\frac{1}{\sqrt{2}}) & (-l_y\tan\theta< y\leq l_y)
\end{cases}
\end{eqnarray}
on the other edge $x=-l_x$, as illustrated in Fig.~\ref{fig:rot_mag_2d} (a).

Figure~\ref{fig:rot_mag_2d} (b) shows the density profile of the zero-energy state in the panel (c).  The zero modes are well separated from 2D bulk bands and are at $(x,y)=\pm(l_x,l_y\tan\theta)$.  In particular, when $\theta=\pm\frac{\pi}{4}$, they are localized at corners and the state is HOTI.  This confirms our construction of 2D HOTI in class AIII by applying symmetry-respecting magnetic fields to 2D $\mathbb{Z}_2$-quantum spin Hall insulator in class DIII.

\begin{figure}[h]
\begin{center}
\includegraphics[clip,width=0.99\columnwidth]{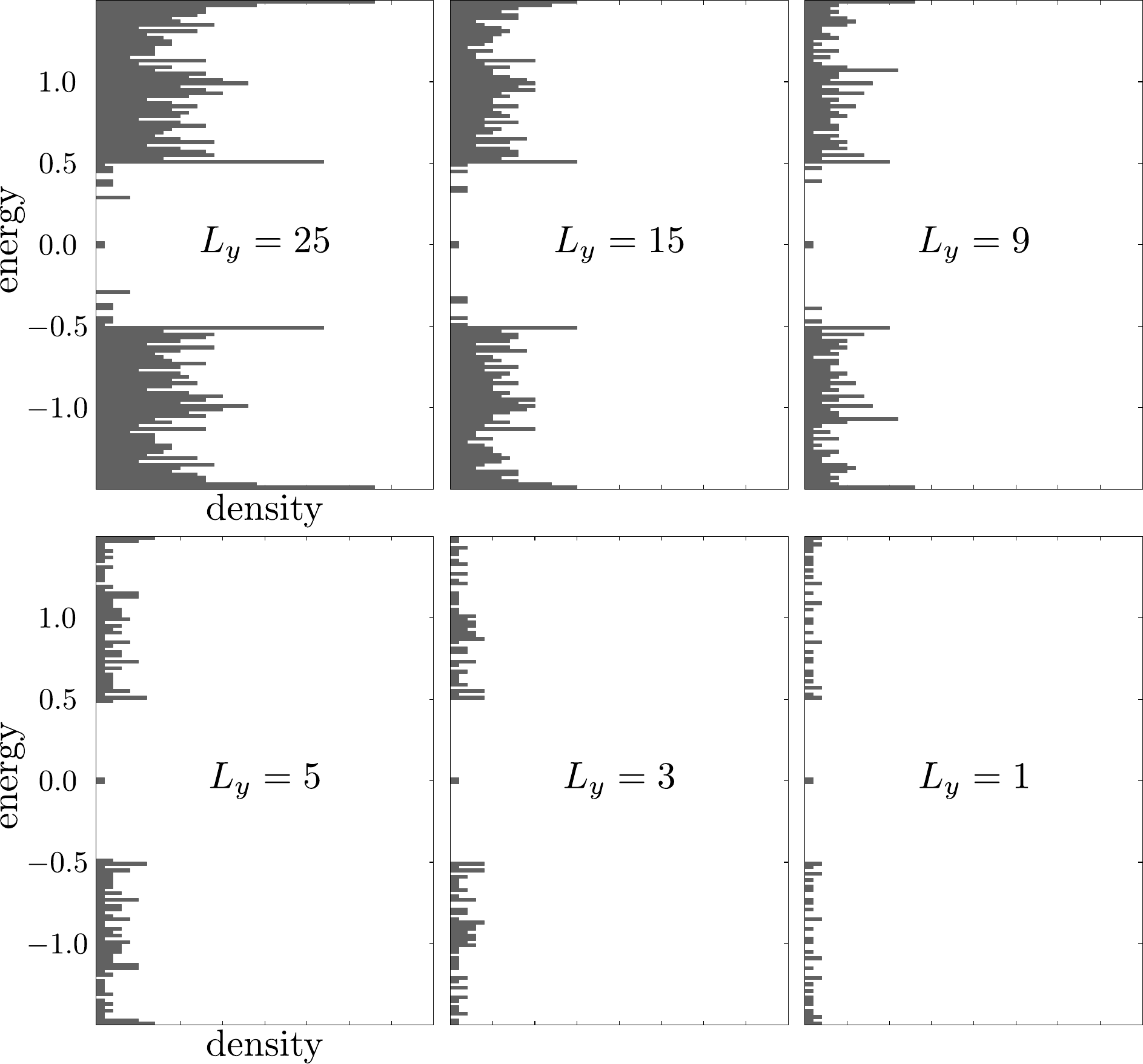}
\caption{The density of states under the OBC in $x$ and $y$ for different number of layers: $L_z=25$, $15$, $9$, $5$, $3$, and $1$.}
\label{fig:2D-tci}
\end{center}
\end{figure}

Now we deform the 2D HOTI to the 1D TI in class AIII.  
We first rotate $\theta$ from $\frac{\pi}{4}$ to $0$. Then reduce the system size in the $y$ direction one by one as in Sec.~\ref{sec:3d-2d}.  Figure~\ref{fig:2D-tci} (d) shows the density of states under the OBC in $x$ and $y$ with reduced layers in the $y$ direction: $L_y=2l_y+1=25, 15, 9, 5, 3, 1$.  The zero modes in the gap remain unaffected in this process; only the bulk density of states is reduced. The resulting 1D insulator is described by the model in Eq.~\eqref{eq:3D-ti-model} with $j=x$ plus the magnetic field $-B_y\tau_0\otimes\sigma_y$.  Therefore, we conclude that the 2D HOTI in class AIII can be smoothly connected 1D TI in the same symmetry class.

\section{Conclusion}
In this paper, we presented two concrete models, one connecting a 3D HOTI to a 2D Chern insulator, and the other relating a 2D HOTI to a 1D TI in class AIII, which exemplify a general understanding of HOTIs in terms of lower-dimensional conventional TIs protected by internal symmetries.  Although our analysis is based on simple specific tight-binding models, it has broader implications because it covers all Hamiltonians of HOTIs that can be smoothly interpolated to our models.

Our 3D HOTI model has a conventional bulk topology (the $\pi$-axion angle) because we started from a $\mathbb{Z}_2$ strong TI and the bulk topology cannot change without closing a bulk gap or breaking the protecting symmetry.  If one wants to do a similar analysis without such a bulk topology, one can prepare a TR copy of our model and form a 
a class AII HOTI model with a 1D helical mode out of them whose the axion angle vanishes.  We can then perform the same analysis independently for the original model and for its TR copy, connecting the class AII HOTI to a 2D quantum spin Hall insulator. In this way, our simple models can serve as building blocks for the discussion of other symmetry classes.

Although HOTIs are a novel class of topological crystalline insulators recently studied extensively, their physical properties may be fully captured by the conventional insulators.  
The HOTI story might still be useful in the material design perspective --- it might give us an easy way of realizing materials with 1D edge states in our 3D space.  Also, it is important to ask if all HOTIs can be constructed by the coupled-layer or coupled-wire type construction or not --- the latter possibility will give us truly new instance of topological crystalline insulators.

\begin{acknowledgments}
HW thank Chen Fang and Hoi Chun Po for useful discussions.  
HW acknowledges support from JSPS KAKENHI Grant Number JP17K17678.
AM acknowledges support from the Materials Education program for the future leaders in Re- search, Industry, and Technology (MERIT).
\end{acknowledgments}

\bibliography{references.bib}

\end{document}